\documentclass[12pt]{iopart}
\usepackage{graphicx}
\usepackage{epsfig}
\usepackage{subfigure}

\usepackage{amsmath}
\usepackage{amssymb}

\usepackage{citesort}

\newcommand{\kHz}{\,\text{kHz}}
\newcommand{\MHz}{\,\text{MHz}}
\newcommand{\erf}{\text{Erf}}
\newcommand{\mm}{\,\text{mm}}
\newcommand{\mV}{\,\text{mV}}
\newcommand{\meV}{\,\text{meV}}
\newcommand{\eV}{\,\text{eV}}
\newcommand{\nm}{\,\text{nm}}
\newcommand{\mW}{\,\text{mW}}
\newcommand{\millis}{\,\text{ms}}
\newcommand{\V}{\,\text{V}}
\newcommand{\mum}{\,\mu\text{m}}
\newcommand{\mus}{\,\mu\text{s}}
\newcommand{\unit}[1]{\,\text{#1}}
\begin{document}

% 32.80.Pj    Optical cooling of atoms; trapping
% 39.30.+w    Spectroscopic techniques
% 03.67.Lx

%\journalname{NJP}

\title{Transport of ions in a segmented linear Paul trap in printed-circuit-board technology}

\author{G. Huber \footnote{corresponding author:
gerhard.huber@uni-ulm.de.}, T. Deuschle, W. Schnitzler, R. Reichle, K. Singer and F.
Schmidt-Kaler}

\address{Institut f\"ur Quanteninformationsverarbeitung, Universit\"at Ulm,
Albert-Einstein-Allee 11, D-89069 Ulm, Germany}

\begin{abstract}
We describe the construction and operation of a segmented linear Paul trap,
fabricated in  printed-circuit-board technology with an electrode segment
width of $500 \mum$. We prove the applicability of this technology to reliable ion
trapping and report the observation of Doppler cooled ion crystals of $^{40}\text{Ca}^+$ with this
kind of traps.
Measured trap frequencies agree with numerical simulations at the level of a few percent
from which we infer a high fabrication accuracy of the segmented trap.
To demonstrate its usefulness and versatility for trapped ion experiments we study
the fast transport of a single ion.
Our experimental results show a success rate of $99.0(1)\,\%$ for a transport distance
of $2\times 2\mm$ in a round-trip time of $T=20 \mus$, which corresponds to 4 axial oscillations only.
We theoretically and experimentally investigate the excitation of oscillations
caused by fast ion transports with error-function voltage ramps: For a slightly slower
transport (a round-trip shuttle within $T=30 \mus$) we observe non-adiabatic motional
excitation of $0.89(15)\meV$.
\end{abstract}
%\pacs{1315, 9440T}

%\tableofcontents

\section{Introduction}
Three-dimensional linear Paul traps are currently used for ion-based quantum
computing \cite{BARRETT2004,RIEBE2004,BECHER2005,WINELAND2005}
and high-precision spectroscopy \cite{SCHMIDT2005,ROOS2006,ROSENBAND2007}. An axial
segmentation of the DC electrodes in these traps enables the combination of individual traps to
a whole array of micro-traps.  Structure sizes down to the range of a few
$10\mum$ \cite{STICK2006,SEIDELIN2006} are nowadays in use
and mainly limited by fabrication technology.
However, most of these fabrication methods require advanced and non-standard techniques
of micro-fabrication as it is known from micro-electro-mechanical system (MEMS) processing or advanced laser machining.
Long turn-around times and high costs are complicating the
progress in ion trap development. In this paper we present a segmented three-dimensional ion trap
which is fabricated in UHV-compatible printed-circuit-board (PCB) technology.
While with this technology spatial dimensions of electrodes are typically limited to more than $100\mum$, the production of
sub-millimeter sized segments is simplified by the commonly used etching process. The advantages
of PCB-traps are therefore a fast and reliable fabrication and consequently a quick
turn-around time, combined with low fabrication costs. 
%REV1
The feasibility of the PCB-technique for trapping ion clouds in a surface-electrode trap has already been shown in~\cite{BROWN2007}.
In future we anticipate enormous impact of the PCB technology by including standard multi-layer techniques in the trap
design. On a longer timescale on-board electronics may be directly included in the layout of the
PCB-boards e.g. digital-to-analog converters may generate axial trap control voltages
or digital RF-synthesizers could be used for dynamical ion confinement.

In this paper we first describe the fabrication of our PCB-trap and the overall
experimental setup in section~\ref{expsetup}. The following section~\ref{trapfrequencies}
is dedicated to a comparison of measured trap frequencies with numerical simulations
for a characterization of radial and axial trapping.
Special attention is payed in section~\ref{coldcrystals} to the
compensation of electrical charging effects of the trap electrodes since
PCB-technology requires insulating grooves between
the electrodes that may limit the performance of the trap. We adjust the
compensation voltages to cancel the effect of stray charges and investigate the stability of this
compensation voltages over a period of months. Our measurements show that PCB-traps are
easy to handle, similar to standard linear traps comprising solely metallic electrodes. In
section~\ref{transport}, the segmentation of the
DC-electrodes is exploited to demonstrate the transport of a single ion,
or a small crystal of ions, along twice the distance of $2\mm$ in a round-trip shuttle over
three axial trap segments. Special attention is given to fast transports within total transport times of T =
$20\mus$ to $100\mus$, corresponding to only $4$ to $20$ times the oscillation period of the ion
in the potential. We checked the consistency of the measured data with a simple transport model.

\section{Experimental setup}\label{expsetup}

\subsection{Design, fabrication and operation of PCB-traps}
\begin{figure}
 \includegraphics[scale=0.55]{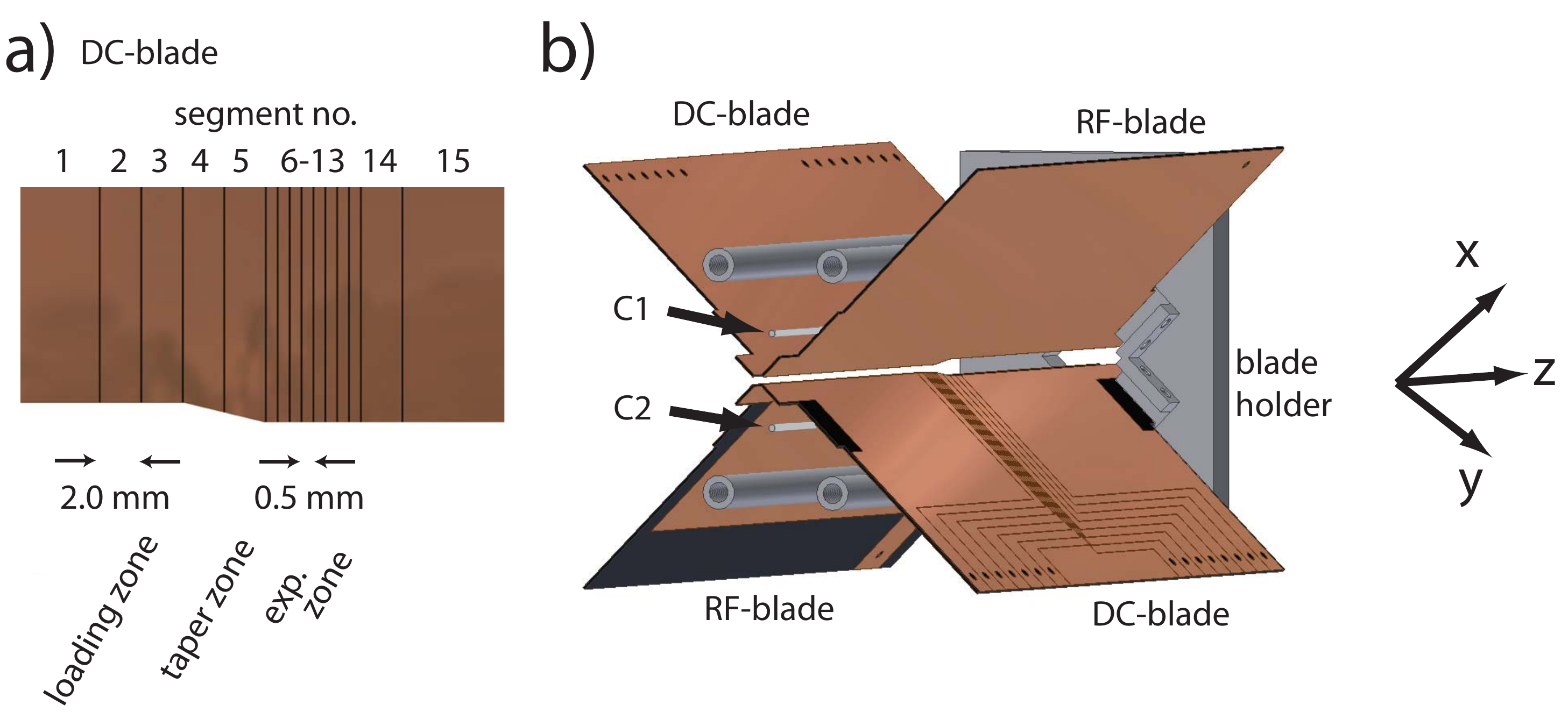} \\
\caption{Electrode design. a) Close-up view of the
blade design: The DC-blade segment width is $2\mm$ in the loading and taper zone, and $0.5\mm$ in the experimental zone.
The trap consists of a $4\mm$ wide loading zone, a tapered intermediate
zone, and a $2\mm$ wide experimental zone. b) Sketch of the assembled X-trap consisting of four blades, drawn with
missing front blade holder for sake of clarity. Compensation electrodes C1 and C2 are parallel
to the trap axis.} \label{fig:schemetrap}
\end{figure}

The trap consists of four blades, two of them are connected to a radio frequency (RF)
supply and the two remaining, segmented blades are supplied with static (DC) voltages,
see fig.~\ref{fig:schemetrap}. The DC- and RF-blades are assembled normal to each other
(the cross-section is X-shaped). In a loading zone the two opposing blades are at a
distance of $4\mm$. A tapered zone is included in order to flatten the potential
during a transport between the wider loading and the narrower experimental zone. In the
latter zone the distance between the blades is reduced to $2\mm$. The material of the
blades is a standard polyimide material coated by $\sim 18\mum$ of copper on all sides of the
substrate. Etched Insulation grooves of $120\mum$ in the copper define DC-electrodes. The
RF-drive frequency near $\Omega/2\pi=11.81 \MHz$ is amplified and
its amplitude is further increased by a helical resonator before applied to the RF
blades. At typical operating conditions we measure a peak-to-peak voltage of about
$V=400 \V_{\rm pp}$ by using a home-built capacitive divider with a small input
capacitance to avoid artificial distortion of the signal by the measurement. The
DC-control voltages from a computer card are connected to the trap segments
via low-pass RC-filters with corner frequency at $1 \MHz$. The trap is housed in a
stainless steel vacuum chamber with enhanced optical access held by an ion
getter- and a titanium sublimation pump at a pressure of 3$\times
10^{-10}\unit{mbar}$. The value was reached without bakeout, indicating the UHV-compatibility of
the pcb-materials \footnote{material P97 by Isola, pcb manufacturer http://www.micro-pcb.ch.}.

\subsection{Laser excitation and fluorescence detection}
\begin{figure}[t]
\begin{minipage}{0.4\linewidth}
\begin{center}
\includegraphics[scale=1]{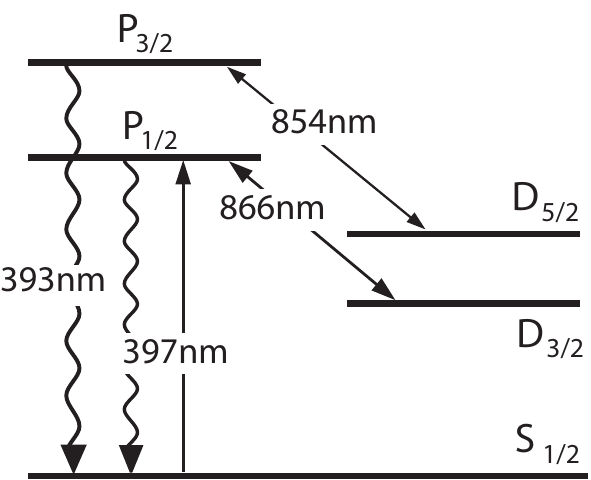}
\end{center}
\end{minipage}\hspace{-1.5cm}
\begin{minipage}{0.65\linewidth}
\caption{Energy levels and relevant optical transitions of $^{40}$Ca$^+$.} \label{fig:levelscheme}
\end{minipage}
\end{figure}

The relevant energy levels of $^{40}$Ca$^+$ ions are shown in
fig.~\ref{fig:levelscheme}. All transitions can be either driven directly by grating
stabilized diode lasers, or by frequency doubled diode lasers \footnote{Toptica DL100, DL-SHG110 and TA100.}. The
lasers are locked according to the Pound-Drever-Hall scheme to Zerodur Fabry-Perot
cavities for long term frequency stability. Each laser can quickly be switched on and off
by acousto-optical modulators. Optical cooling and detection of resonance fluorescence
is achieved by simultaneous application of laser light at $397 \nm$ and $866 \nm$.
Radiation near $854\nm$ depopulates the D$_{5/2}$ metastable state
($1.2\unit{s}$ lifetime).

Two laser sources at $423 \nm$ ($\le 5 \mW$) and $375 \nm$ ($\le 1 \mW$) are used for
photoionization loading of ions \cite{GULDE01,KIER00}. The high loading efficiency allows for a
significant reduction of the Ca flux from the resistively heated oven and minimizes the
contamination of the trap electrodes. By avoiding the use of electron impact ionization the occurrence of stray charges
is greatly suppressed.

In this experiment laser beams near $397 \nm$, $866 \nm$ and $854\nm$ are superimposed and
focused into the trap along two directions: one of these directions (D1) is aligned vertically under $5^\circ$
with respect to the top-bottom axis and the other beam (D2) enters horizontally and 
intersects the trap axis under an angle of $45^\circ$. A high-NA lens is placed $60\mm$ from the trap center behind an inverted viewport to monitor the fluorescence of trapped ions at an angle perpendicular to the trap axis direction.
The fluorescence light near $397\nm$ is imaged onto the chip of an EMCCD camera \footnote{Electron Multiplying CCD camera, Andor iXon DV860-BI.}
in order to achieve a photon detection efficiency of about 40\,\%.

\section{Numerical simulations }\label{trapfrequencies}
\subsection{Radial confinement}
\begin{figure}[htpb] %\resizebox{0.50\hsize}{!}
\centering
\includegraphics[width=0.85\textwidth]{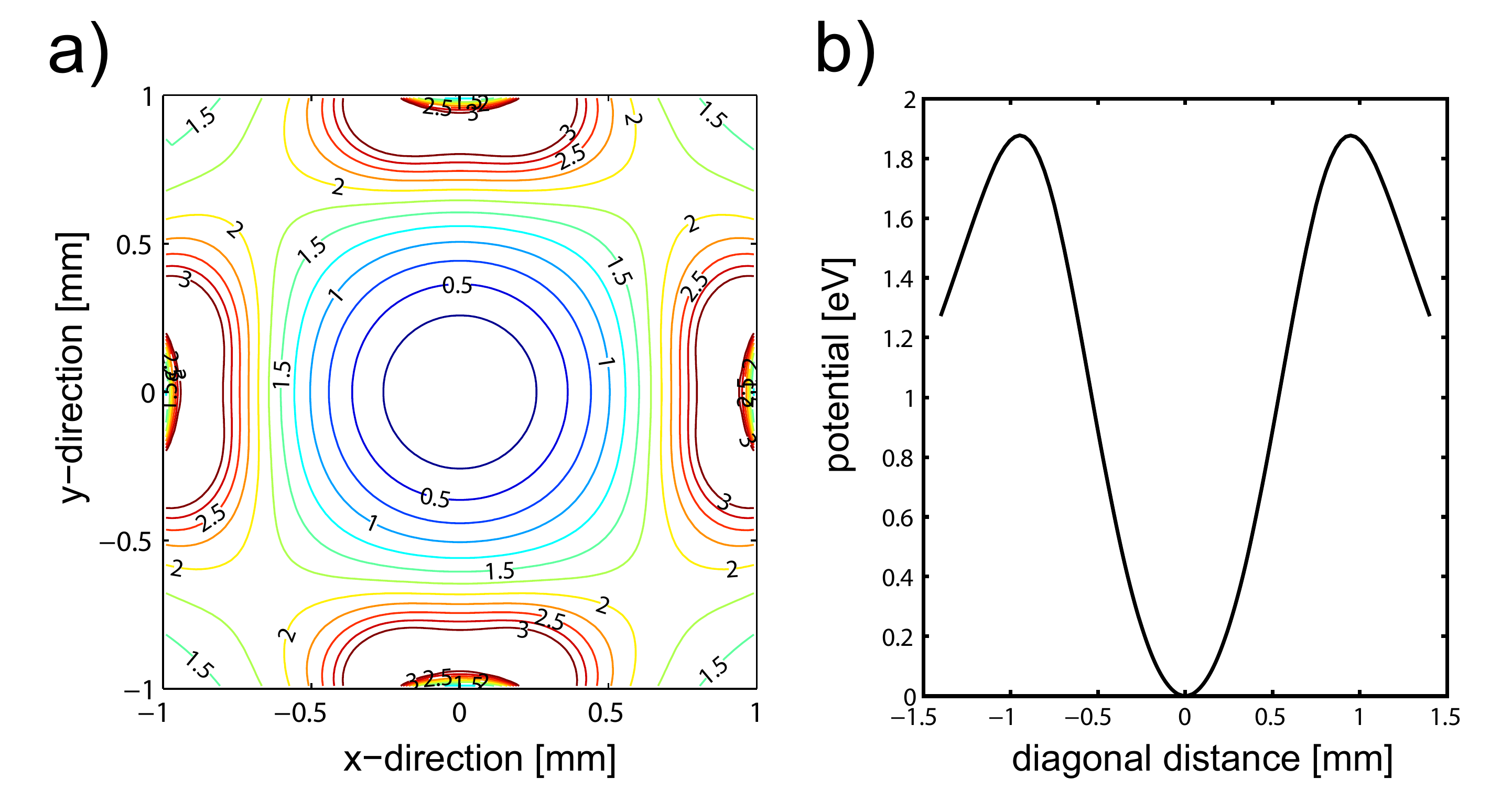}
\caption{a) Simulation of the pseudo-potential in the x-y plane in the experimental zone for
$V=408\V_{\rm pp}$. b) Cross-section through the potential shown in a) along the $x=y$ direction
yielding a radial trap frequency of $\omega_{\rm rad}/2\pi= 663 \kHz$ and a trapping depth of approx. $1.9\eV$.} \label{fig:potradial}
\end{figure}

%REV1
The complex layout of the multi-segmented trap requires elaborate numerical techniques
to simulate the electrical potentials accurately in order to avoid artificial effects. Instead of
using widespread commercial FEM routines that mesh the whole volume between the electrodes,
boundary-element methods (BEM) are more suited but also more complex in their
implementation. The effort in such methods is reduced for large-space computations
since they only need to mesh surfaces of nearby electrodes and multipole approximations
are done to include far distant segments in order to speed up computations. We have written a framework 
for the solution of multipole-accelerated Laplace problems\footnote{The solver was developed by the MIT Computational Prototyping Group, http://www.rle.mit.edu/cpg.}. This method is not limited in the
number of meshed areas and thus allows for accurate calculations with a fine mesh.
In the simulations presented here, we typically subdivide the surfaces of the trap
electrodes into approx. 29000 plane areas to determine the charge distributions and the
free-space potential within the trap volume numerically. Fig.~\ref{fig:potradial}a
shows the calculated pseudo-potential in the x-y-plane of the experimental zone for a
radio frequency of $\Omega/2\pi=11.81 \MHz$ and a RF peak-to-peak voltage $V=408 \V_{\rm
pp}$. By symmetry the diagonal directions along $x=\pm y$ contain the nearest local
maximum and thus determine the relevant trap depth of approx. $1.9\eV$ for the
applied experimental parameters. A cross-section through the pseudo-potential along
this direction is shown in fig.~\ref{fig:potradial}b at the same axial position as in fig.~\ref{fig:potradial}a. From
these simulations we are able to extract a radial trap frequency of $\omega_{\rm
rad}/2\pi= 663 \kHz$ in the absence of applied DC potentials.

\subsection{Axial confinement}\label{sec:axialconfinement}

\begin{figure}[t]
\begin{minipage}{0.7\linewidth}
\begin{center}
\includegraphics[width=0.8\textwidth]{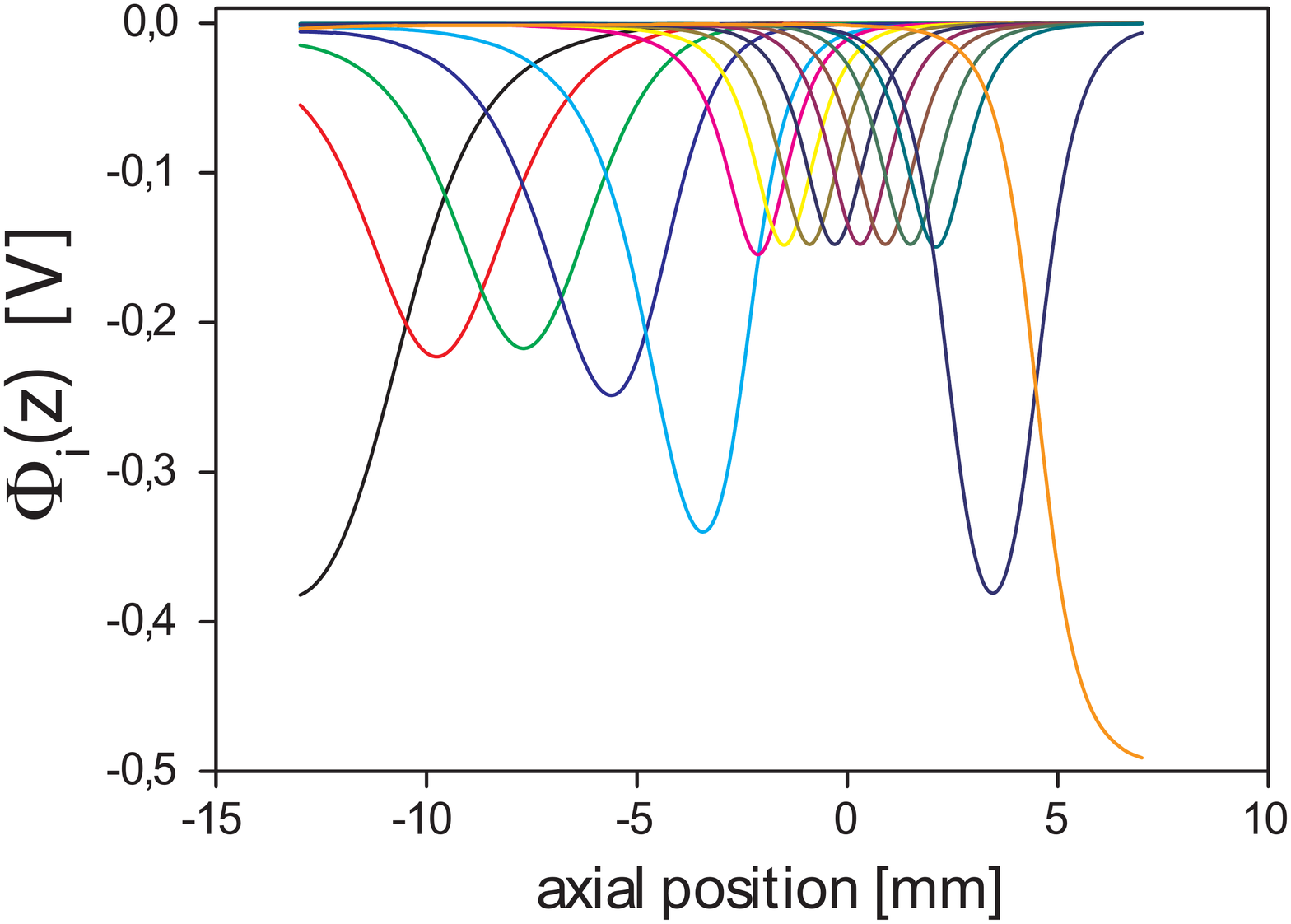}
\end{center}
\end{minipage}\hspace{-3cm}
\begin{minipage}{0.45\linewidth}
\caption{Individual axial electrode pair potentials $\phi_i(z)$. For detailed description see text in section~\ref{sec:axialconfinement}.
Narrow electrodes show a lower maximum potential due to the smaller amount of surface charges.
An arbitrary axial potential is formed by superposing individual contributions.}
\label{fig:potaxial}
\end{minipage}
\end{figure}

For axial confinement opposing DC electrodes are set to the same voltage $U_i$, $i=1,...,15$ labeling the electrode number as depicted in fig.~\ref{fig:schemetrap}. These voltages are supplied by digital-to-analog (DA)
converters covering a voltage range of $\pm 10\V$ with a maximum update rate of $1\MHz$.
Fig.~\ref{fig:potaxial} illustrates the potentials $\phi_i(z)$ obtained when only $U_i$ is set to $-1 \V$ and all other voltages are set to zero.
For arbitrary voltage configurations the axial potential $\phi(z)=\sum_i U_i\phi_i(z)$ is a linear superposition of the single electrode potentials. Tailored axial potentials for the generation of single or multiple axial wells with given
secular frequency, depth and position of the potential minimum are obtained by solving the linear equation given above for the voltages $U_i$. Time dependent voltages are used to move the potential minimum guiding the trapped ion along the trap axis.

\section{Cold ion crystals} \label{coldcrystals}%daten vom 29.1.2007
\subsection{Observation of linear crystals and measurement of trap frequencies}
The trap is operated in the loading configuration with DC voltages of
$ U_\text{load}=\{\ldots, U_7,\ldots U_{13}, \ldots\}_{t=0} =\{ \ldots, 6 \V, \ldots, 8 \V, \ldots\}$
while non-specified segment voltages are held at ground potential.  From simulations we estimate that
the axial trap frequency is $\omega_{z}/2\pi= 199 \kHz$ at the potential minimum close to
the center of electrode $10$.
Linear strings and single ions crystallize under continuous Doppler cooling
with beams D1 and D2, consisting of superimposed beams at $397\nm$ ($0.3\mW$ with $60\mum$ beam waist), 
$866\nm$ and $854\nm$ (both $3\mW$ with $100\mum$ beam waist).

Applying a sinusoidal waveform with $1\mV$ amplitude to electrode segment $10$ allows for
a parametrical excitation near resonance.
Observation of the ion excitation through decreasing
fluorescence on the EMCCD image signifies a resonance condition.
This way, we experimentally find a radial frequency of $\omega_{rad}/2\pi=700(2) \kHz$
and an axial frequency of $\omega_{z}/2\pi=191(2) \kHz$. We
attribute the $5\,\%$ deviations to residual electric fields arising mainly from charging
of the etched insulation grooves cuts as well as to the limited accuracy of the RF-voltage amplitude
measurement. Stable trapping at a $q$-value of about $0.16$ is achieved.
Measuring the mutual distances in a linear two- or three-ion crystal easily allows for
calibrating the optical magnification of the imaging optics to be $10.4$, with a CCD-pixel size of $24\mum$.
For determining the resonant frequency of the cycling transition used for  Doppler cooling the fluorescence is detected on a photomultiplier tube (PMT) while scanning the
laser frequency at $397 \nm$ and/or $866 \nm$. This yields an asymmetric line profile of
about $30-50 \MHz$ FWHM, exhibiting the features of a dark resonance with a maximum
count rate of $20 \kHz$. For all following measurements we keep
the detuning of the $397 \nm$ laser at $\Gamma/2$ (half of the linewidth) fixed and adjust the frequency of
the $866 \nm$ laser to obtain a maximum fluorescence rate.

\begin{figure}[t]
\begin{minipage}{0.5\linewidth}
\begin{center}
\includegraphics[width=1.0\textwidth]{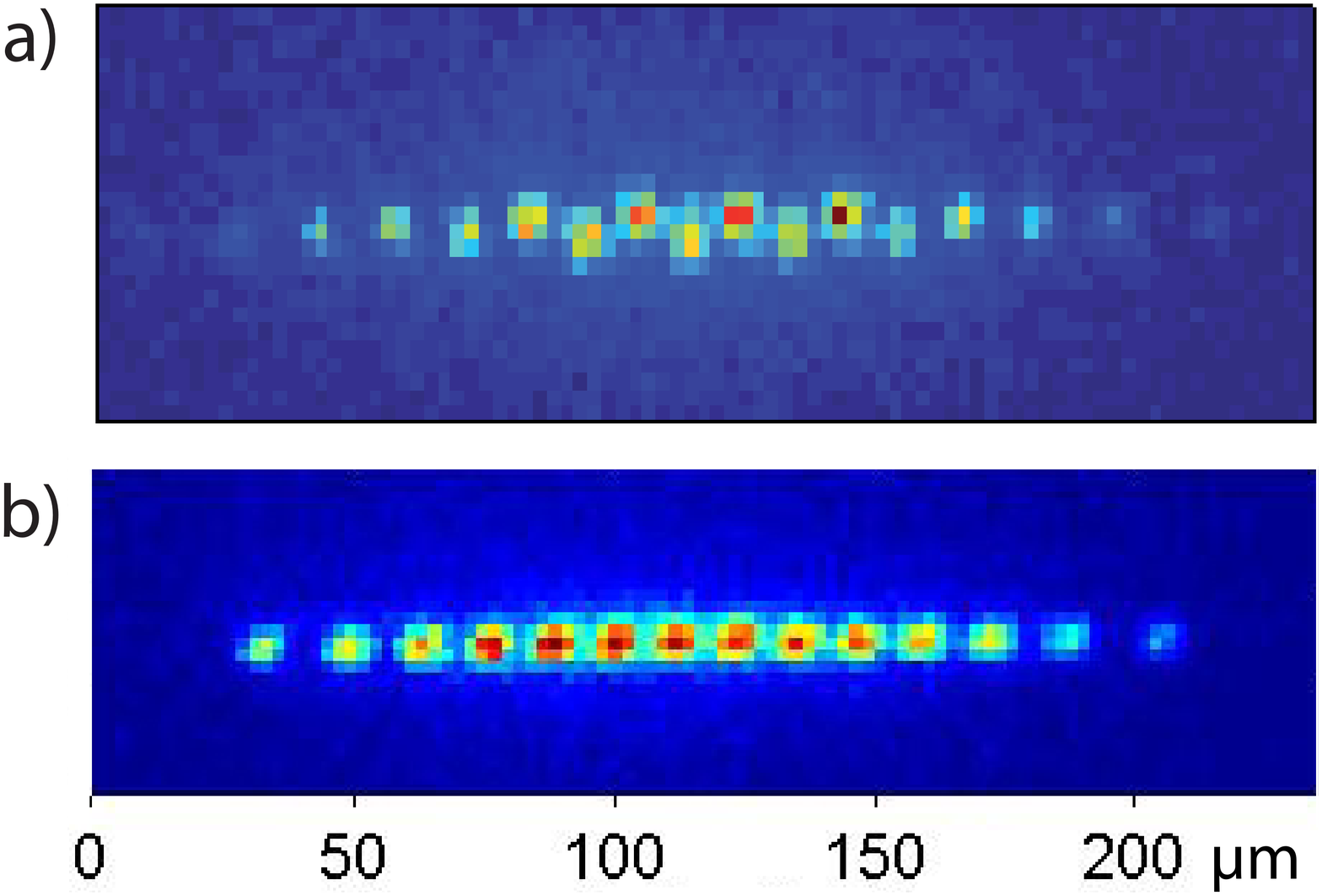}
\end{center}
\end{minipage}\hspace{-1.5cm}
\begin{minipage}{0.5\linewidth}
\caption{a) Fluorescence of an ion crystal with 14 ions in a zig-zag configuration
as observed with the EMCCD camera. Exposure time was set to $100 \millis$ (optical magnification is $10.4$).
b) Linear crystal with 14 ions and a slightly more relaxed axial trap frequency than in a).
This induces a phase transition into a purely linear configuration of ion positions.} \label{fig:kristalle}
\end{minipage}
\end{figure}

\subsection{Compensation of micro-motion} \label{compensation}
Ions in small traps are likely to be affected by stray charges that shift the ions out
of the RF null since those charges can often not be neglected in most ion traps.
Therefore, the dynamics of ions in general exhibits a driven micromotion oscillation
at a frequency of $\Omega/2\pi$ leading to a broadening of the Doppler
cooling transition and higher motional excitation \cite{BERKELAND}. A second
disadvantage of micromotion is a reduced photon scattering rate at the
ideal laser detuning of $\Gamma/2$.
To correct for potentials induced by stray charges and maybe geometric
imperfections we can apply compensation voltages to the electrodes labeled
C1 and C2, as sketched in fig.~\ref{fig:schemetrap}b.

\begin{figure}[htpb] \centering \includegraphics[width=\textwidth]{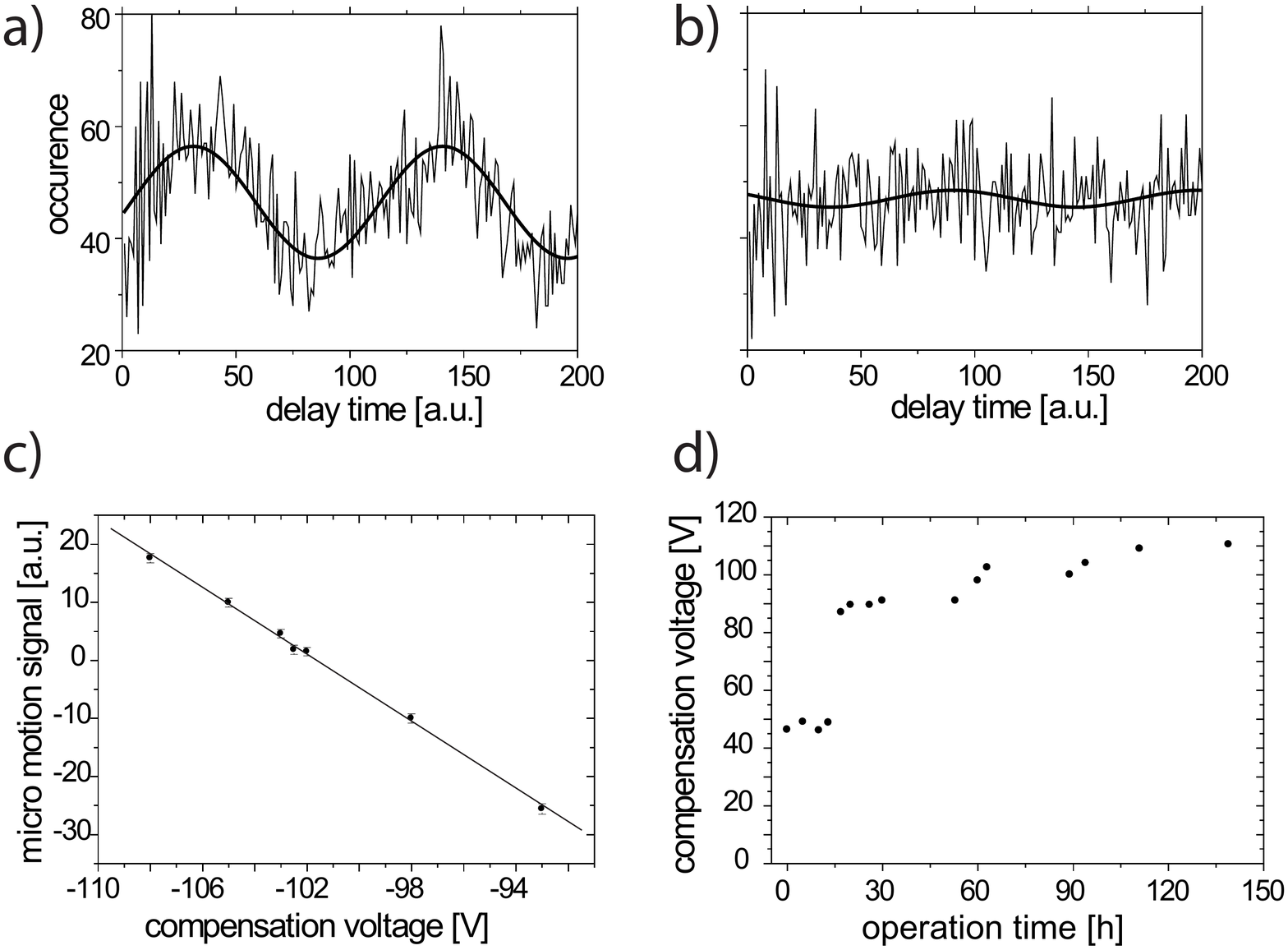}
\caption{Histogram of photon count events for different values of micromotion
compensation voltages: a) trace far from compensation with $U_\text{C1}=98 \V$
fitted to a sine function.
b) near to optimum compensation with $U_\text{C1}=102.5\V$. c) The amplitude of a
sinusoidal fit for different compensation voltages shows a linear dependence
between the amplitude and applied compensation. 
%REV1
Negative amplitudes correspond to a $180^\circ$ phase flip; this happens when the ion passes the RF null.
A linear fit reveals the optimal value at $U_\text{C1}=101.6 \V$. With this data an optimal compensation is
determined within an uncertainty of less than $0.3\V$. The overall data collection time is less than 5\,min.
d) Optimum compensation voltage plotted versus the operation time of the oven.
We attribute the long term drift of this voltage to an increasing calcium contamination
on the trap electrodes in the experimental zone.} \label{fig:komp}
\end{figure}

%REV1
For detecting the micromotion we trigger a counter with a
photon event measured by the PMT in the experimental zone and stop the counter by a TTL phase locked to half the
trap drive frequency $\Omega/2\pi$. If the ion undergoes driven motion, the
fluorescence rate is modulated as a cause of a modulated Doppler line shape, see
fig.~\ref{fig:komp}a. Here, we detect the ion motion in the direction of the laser beam.
If a flat histogram is measured the ion does not possess a correlated motion with the
RF driving voltage. Then, it resides close to the RF null where the modulation vanishes, cf.
fig.~\ref{fig:komp}b. From taking a series of histograms for different values of the compensation
voltage a linear relation between observed modulation and applied compensation is obtained.
%REV1
The optimal compensation voltage can then be read from the abscissa for zero correlation amplitude, fig.~\ref{fig:komp}c.
\par
Our long term observation of the optimum compensation voltage indicates a weak increase
over three months.
Since we did not bakeout the vacuum chamber, storage times of ions are
several minutes so that we needed to reload ions from time to time.
During all experimental sessions we kept the temperature of the
oven fixed so that the flux of neutral Ca atoms was sustained. This is not necessary in a better vacuum environment.
An oven current of $7\unit{A}$  leads to a flux such
that we reach a high loading rate of $0.1$ to $0.4$ ions per second.
Thus, by operating the oven continuously we could accumulate a long operation time of over $150\unit{h}$.
The $20\,\%$ step in fig.~\ref{fig:komp}d was caused by a power failure.

By comparing our measurements to the long
term recording of Ref. \cite{ROTTER2003} for a $2\mm$ linear trap \cite{SK2003}
with stainless steel electrodes we find similar drifts in compensation voltages.
Thus, it seems that PCB-traps can be corrected equally well for micromotion so that
the larger insulation grooves do not harm the performance.

In all of the experiments presented here the beam of neutral calcium is directed into the narrow
experimental zone where micromotion is compensated as described above.
In future, however, we plan to use the loading zone where a higher
loading rate through the larger involved phase space during trapping is expected.
Then, ions can be shuttled into the cleaner experimental zone $8\mm$ apart.
Calcium contamination will then be completely avoided and much lower
compensation voltages and drifts are expected.
The segmented PCB-trap operated this way
may even show an improved micromotion compensation stability as compared to traditional
linear and three dimensional traps having less insulation exposed to the ions.

 % daten vom 23.4.7

\section{Transport of single ions and ion crystals \label{transport}}
As recently suggested \cite{LEIBFRIED2007}, ions in a future quantum computer based on
a segmented linear Paul trap might be even shuttled through stationary laser beams to
enable gate interactions. In a different approach one might shuttle ions between the quantum logic operations \cite{KIELPINSKI2002}. This would alleviate more complex algorithms and reduce the enormous technological requirements extensively. According to theoretical investigations \cite{ChuangPC}, transport could occupy up to $95\,\%$ of operation time in realistic algorithms. Given the limited coherence
time for qubits, shuttling times need therefore to be minimized. 
Our work as described below may be seen in the context of theoretical and experimental work found in \cite{HOME03, HUCUL07} and \cite{HENSINGER06}.
With our multi-segmented Paul trap we have made first investigations of these speed limits in order to enter
the non-adiabatic regime and studied the corresponding energy transfer. Preferred 
transport ramps are theoretically well understood \cite{REICHLE2006}. The current challenge arises from 
experimental issues, i.e. how accurate potentials can be supplied, how drifts in
potentials can be avoided and whether sophisticated shuttling protocols can be used
\cite{SCHULZ2006}. First transport studies have been made some years ago by the Boulder
group using an extremely sparse electrode array \cite{ROWE2002}. Non-adiabatic transports of cold neutral atom clouds in a dipole trap have been demonstrated in~\cite{COUVERT07} and show very similar qualitative behavior, even though on a completely different time scale (axial frequency of approx. $2\pi\times 8\unit{Hz}$).

The PCB-trap described here contains 15 pairs of DC-segments. Using a suited time sequence of control
voltages $U_i(t),\ i=1,...,15$, the axial potential can be shaped time-dependently in
a way to transport ions along the trap axis. In the following we present
the results of various transport functions for a distance of $2d=4\mm$.

\begin{figure}
\hspace{-1cm}
\begin{minipage}{0.7\linewidth}
\includegraphics[width=\textwidth]{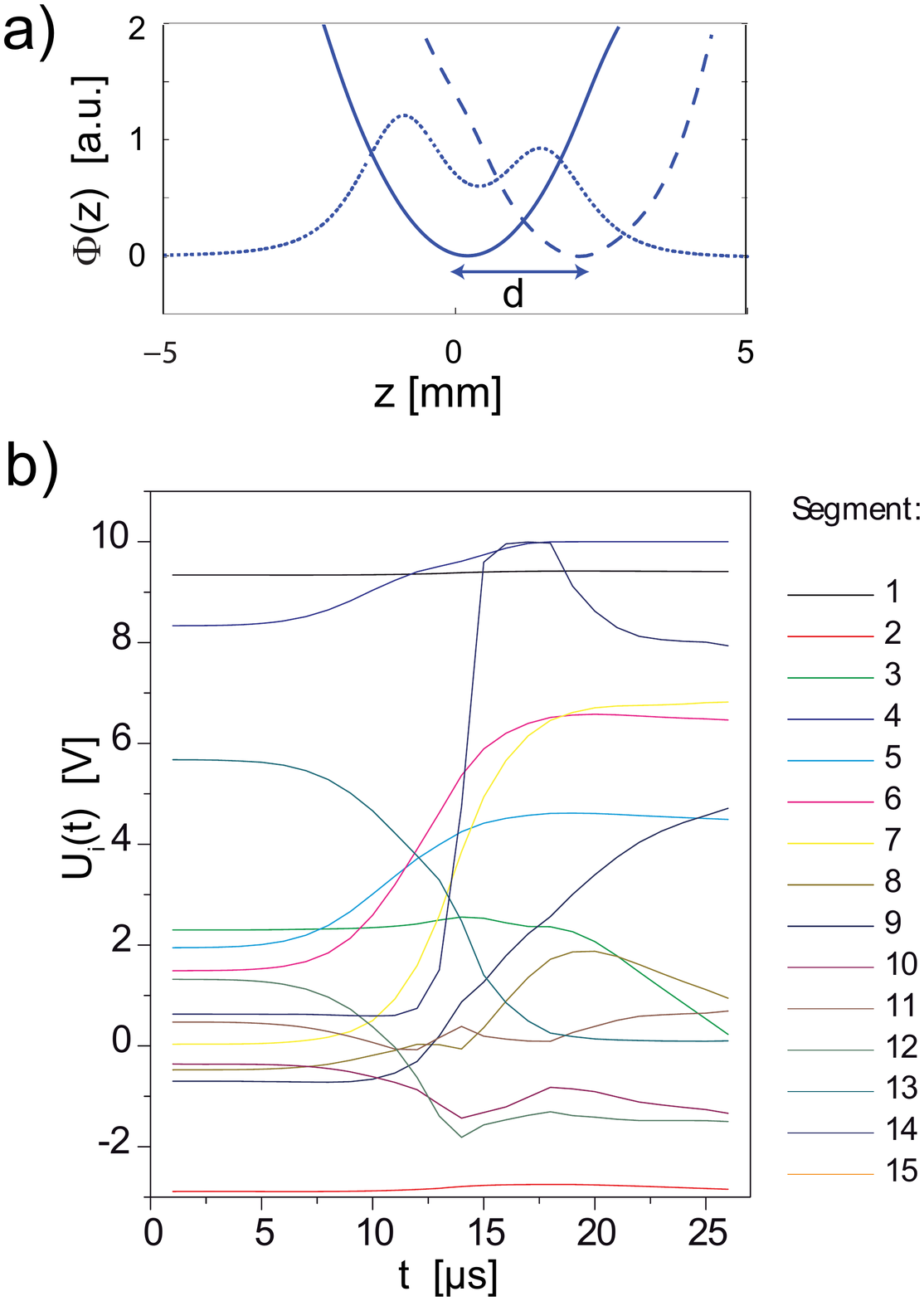} 
\end{minipage}\hspace{-1.5cm}
\begin{minipage}{0.5\linewidth}
\caption{a) Axial potentials for loading and observing ions (dotted line) $U_\text{load}$,
at the beginning (solid line) of the transport with
\begin{align*} &U_\text{trans}(t=0) = \\
& \{-8.77, 9.34, -2.89, 2.30, 8.33, \\
&  1.95, 1.49, 0.03,-0.48,-0.70, \\
&  -0.36, 0.47, 1.32, 5.68, 0.63\}\V\end{align*} and at its turning
point (dashed line). b) Trap control voltages $U_1$ to $U_{15}$ for transporting the ion by
an error function ramp. Discontinuities in the curves result from the time
discretization in steps of 1 $\mus$.}
\label{fig:loadtransport}
\end{minipage}
\end{figure}

For these measurements we follow a 6-step sequence:
\begin{description}
    \item [i)] Initially, we confine an ion in
an axial trapping potential in the loading configuration $U_\text{load}$, fig.~\ref{fig:loadtransport}a. The single ion is
laser cooled by radiation near $397 \nm$ and $866 \nm$.  Compensation voltages have previously been
optimized for this trap configuration such that the width of the
excitation resonance near $397 \nm$ is minimal.  Then, the laser beams are switched off.
 \item [ii)] Before starting the transport, $U_\text{load}$ is linearly changed into the initial
transport potential $U_\text{trans}$(t=0) in 10 steps each taking $1 \mus$. Note, that the transport potential should be adapted in shape and depth, and is not necessarily identical to the optimum potential for laser excitation, for fluorescence observation, and for quantum logic gate operations. In our case, the axial trap frequency for $U_\text{trans}$(t=0) is adjusted for
$\omega_{\rm z}/2\pi = 200 \kHz$, and the minimum positions of $U_\text{load}$ and
$U_\text{trans}$ both coincide with the center of segment 10. The voltages for the
transport potential are chosen according to the regularization of Ref. \cite{REICHLE2006}.
\item [iii)] By changing the
control voltages such that the potential minimum of $U_\text{trans}$ moves according to an
error function, the ion is accelerated and moved to a turning point $2\mm$ away,
centered roughly above segment 13.
\item [iv)]
The ion is accelerated back to the starting point again using the same time-inverted
waveforms. By our calculations we aim to determine the control voltages such that the
ion always remains within a harmonic potential well of constant frequency $\omega_\text{z}$
throughout the whole transport procedure.
\item [v)] Finally, the transport potential $U_\text{trans}$ is ramped back linearly in $10 \mus$
into the initial potential $U_\text{load}$.
\item [vi)] The laser radiation is switched on
again to investigate either the success probability of the transport or the motional
excitation of the ion.
\end{description}
The sequence is repeated about $10^3$ to $10^5$ times, then
the parameters of the transport ramp in step (iii) and (iv) are changed and the
scheme is repeated.

Ideally, we would create an axial potential that closely approximates
$\phi(z, t) = m \omega_{\rm z}^2 (z-z_0(t))^2/2q$
with an explicit time dependent position of the potential minimum $z_0(t)$.
However, all presented simulations and motional excitation energies were deduced from
the numerical axial potentials instead. Moving the potential minimum from its starting
point $z_0(0)=0$ to the turning point $z_0(T/2)=2\mm$
and back again to $z_0(T)=0$ leads to a drag on the trapped ion. For sake of convenience, we express
the total time for a round-trip $T$ in units of the trap period $\tau\equiv T \omega_{\rm
z}/2\pi$. The dimensionless parameter $\tau$ then denotes the number of oscillations
the ion undergoes in the harmonic well during the transport time. The functional form
of the time dependent position of the potential minimum $z_0(t)$ is crucial for the
transport success and the motional excitation. We have chosen a truncated
error-function according to
\begin{equation}
f_{\sigma}(t) = \frac{d}{2}\left(1+\frac{\erf\left[\left(4t/T-1\right)\sigma\right]}{\erf[\sigma]}\right), \quad
z_0(t)=\left\{\begin{matrix} f_{\sigma}(t), & 0\le t \le T/2 \\ d-f_{\sigma}(T-t), & T/2<t\le T \end{matrix}\right.
%\begin{align} 
\label{eq:erf}
\end{equation}
%\end{align}
as input to find the waveforms of all contributing electrode pairs,
fig.~\ref{fig:loadtransport}b.

\begin{figure}
\centering
\includegraphics[width=1.0\textwidth]{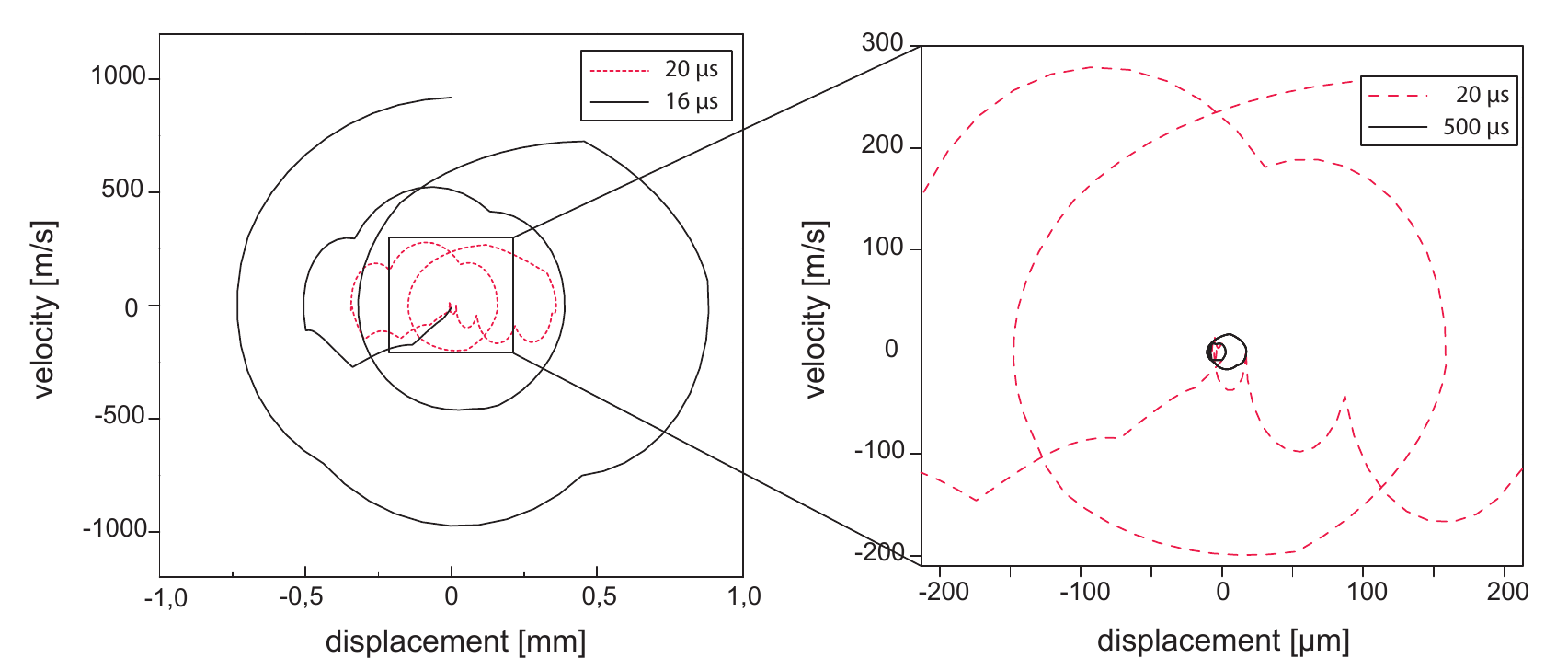}
\caption{Phase space trajectories of an ion for two different transport durations. The
closed loop (dotted line, $\tau=4$) indicates a transport with low final motional energy.
For $\tau=3.2$ (dashed line) the ion does not reach zero
velocity at $t=T$ and a final energy of $171\meV$. Discontinuities in the trajectories
result from the time discretization in steps of $1 \mus$.}\label{fig:phasetrajectory}
\end{figure}

Small first and second derivatives at the corner points
assure that the ion experiences only smoothly varying accelerations at these times. The
parameter $\sigma$ determines the maximum slope of the function. The smoother the shuttling
begins the higher average and maximum velocities are needed for a fast transport.
A high maximum velocity $\dot{z}_0$ results in far excursions of the ion from the
potential minimum so that it may experience higher derivatives not fulfilling the harmonic
potential approximation anymore. 
A compromise can be found between slowly varying corner point
conditions and ion excitation due to fast, non-adiabatic potential movements.
Experimentally, we found the lowest energy accumulated for $\sigma=2.3$.
In our experiment, the update time of the supply hardware for the electrode voltages is limited to
$1 \mus$. To account for this effect in our simulations we also discretize $z_0(t)$ in time steps.
These discontinuities are evident in fig.~\ref{fig:loadtransport}b already indicating
that a higher sampling rate of the DA channels would be desirable.
%REV1
Furthermore, for short shuttles the amplitude discretization through
the DA converters makes an exact reproduction of $z_0(t)$ impossible.
This gives rise to discrete dragging forces transferring motional excess energy to the ion.
We verified by numerical simulations that even for
the fastest transports with durations of only $T\approx 10-12 \mus$, deviations from the non-discretized time evolution were negligible.
Since in our case the typical involved motional energies are much larger than $\hbar\omega$, a simple classical and one-dimensional model of the ion transport is justified with the equation of motion
\begin{equation}
 \ddot{z}(t) = -\frac{q}{m} \frac{d\phi(z,t)}{dz} = -\omega^2 (z-z_0(t))^2.
\end{equation}
We solve this equation numerically for functions $z_0(t)$ with varying $\sigma$. The
resulting phase space trajectories $\{z(t), \dot{z}(t)\}$ are plotted in
fig.~\ref{fig:phasetrajectory} for $T=20 \mus$ ($\tau=4$) and $T=16\mus$ ($\tau=3.2$).
In the first case, the phase space trajectory starts and ends close to
$\{$0,0$\}$, i.e. both the potential and the kinetic energy are modest after the transport.
In the second case, the particle reaches its starting point $z_0(T)=z(0)$ again,
but resides with a high  oscillating velocity $\dot{z}(T)$ but small displacement.

\subsection{Transport probabilities}
This section addresses the question of how fast the transport of an ion in the
manner described above can eventually be performed before it gets lost. In the
following we discuss transports in the adiabatic regime $\tau\gg 1$ and far off this regime. For
the measurements, an experimental sequence with transport duration $\tau$ is interleaved by Doppler
cooling cycles of $1\unit{s}$ duration to ensure to start always from the same initial
energy close to the Doppler limit. After $n_i$ successful transports with $i=1,...,N$ the
ion may finally be lost. After  $100-1000$ repetitions we bin the data into a histogram of
$n_i$ approximating the success probability $P_\tau(n)$ that allows for deduction of the
fraction of ions having performed at least $n$ successful transports. $P_\tau(n)$ is
fitted to an exponential decay introducing the single transport success probability 
$\widetilde{p}_\tau$, $P_\tau(n)=(\widetilde{p}_\tau)^n$. The probability for $n$ successful transports equals
$(\widetilde{p}_\tau)^n$ as the processes are independent with
a sufficiently long Doppler cooling period. To account for other sources of ion loss,
e.g. from background gas collisions, a loss rate without transport in sequence steps
(iii) and (iv) was subtracted. These losses can be modeled by introducing a second
decay channel in $P_\tau(n)$ to finally yield net transport probabilities $p_\tau$.

\begin{figure}[htpb] \centering
\includegraphics[width=0.6\textwidth]{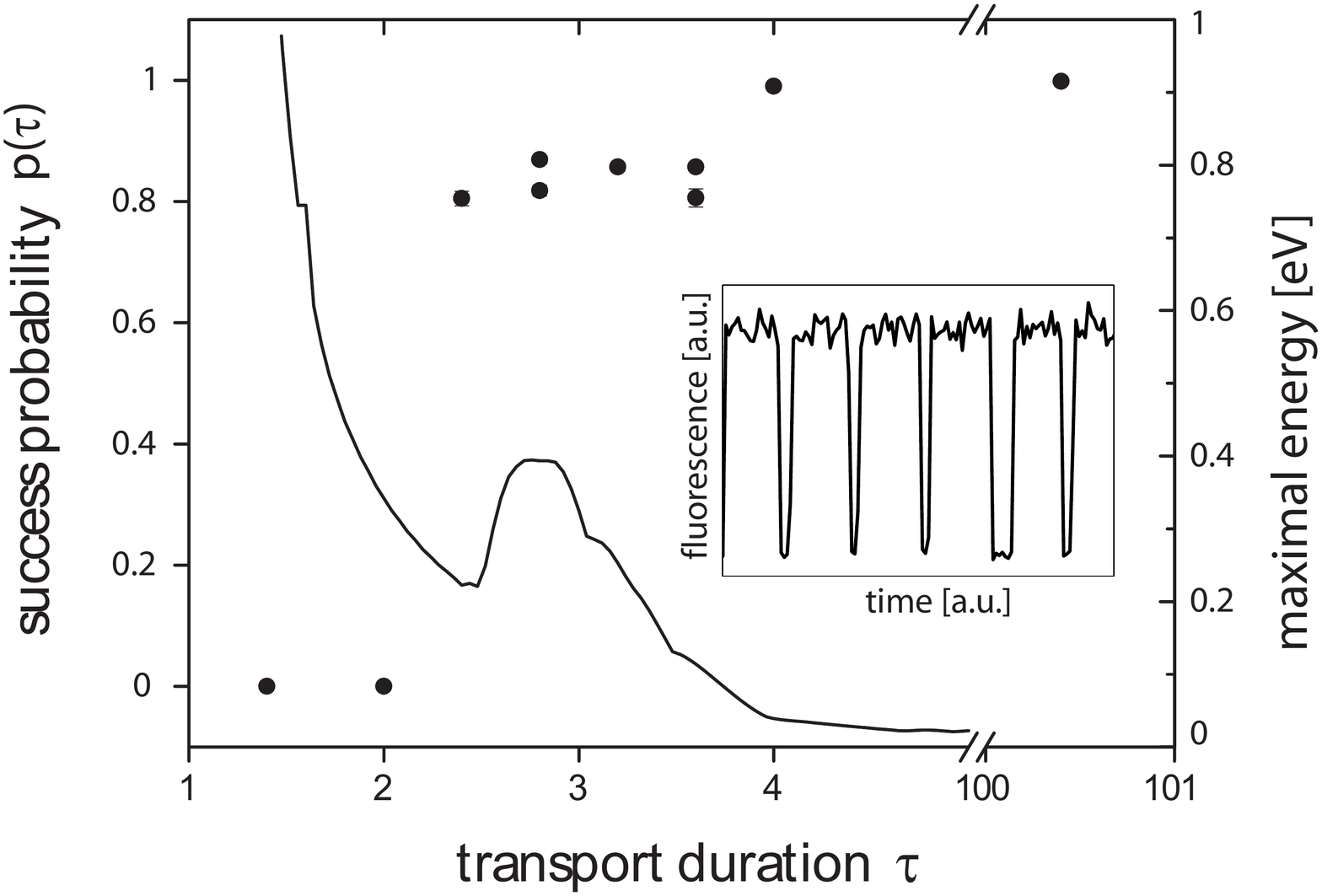}
\caption{Transport success probability as a function of transport time (dots),
y-scale on left. Slow transports with large $\tau$ show a success probability close to
unity. In the intermediate range $2.5 \le \tau < 4$, the probability reduces to approx.
$85\,\%$. From the model we deduce the ion's maximum energy during the transport (solid),
y-scale on right side. Inset: Fluorescence level as observed by the CCD camera showing the
dark periods during the transport while the bright periods indicate a successful
transport.} \label{fig:successprob}
\end{figure}

Fig.~\ref{fig:successprob} depicts values $p_\tau$ for different transport durations $\tau$ performed with a transport function $z_0(t)$ according to equation (\ref{eq:erf}) with $\sigma=2$. In the adiabatic regime we obtain a success probability of 99.8\,\% for
$\tau=100$. The ion stays deep within the potential well guaranteeing low losses. This
is in good agreement with theoretical predictions. The very high success
probability stays almost constant within the adiabatic regime down to $\tau=4$ ($p_4=99.0\,\%$). According to our model the ion experiences a relative displacement of over
$300\mum$ gaining a potential energy of more than $30\meV$. Even faster transports lead
to higher transport losses resulting in probabilities around $85\,\%$. When the energy of the
ion exceeds about $30\,\%$ of the depth of the potential, we observe a strongly increased loss probability.

\subsection{Coherent excitation during transport}
If the ions are to be laser-cooled after a fast shuttle, it is important to keep the
excitation of vibrational degrees of freedom minimal. Therefore, we quantitatively investigate the ion's
kinetic energy after the transport for ramps in the
non-adiabatic regime with $4<\tau\le20$. We generalize a method which was recently employed
\cite{WESENBERG2007,REICHLE200x} to measure motional heating rates in a micro ion trap.

\begin{figure}[htpb]
\begin{minipage}{0.6\linewidth}
\includegraphics[width=\textwidth]{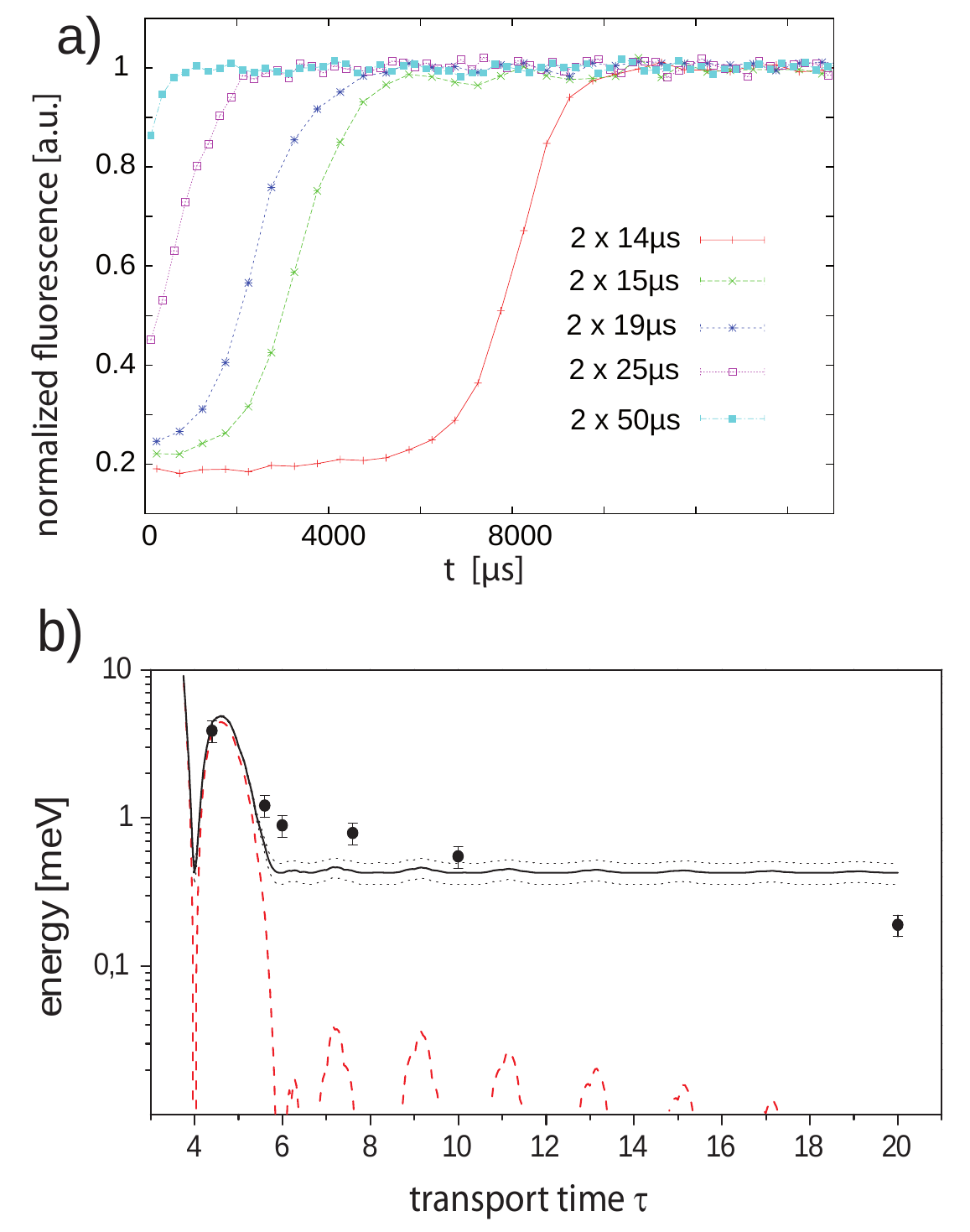}
\end{minipage}\hspace{-1.5cm}
\begin{minipage}{0.5\linewidth}
\caption{a) Fluorescence rate after the transport and during application of Doppler
cooling. Fast transport results in large excitation and a late recovery of the
fluorescence level from vibrational excitation.
b) Measured excitation energy after an error function transport
with $\tau = $~4 to 20. Error bars of the data points account for the uncertainty of
the excitation energy from the photon scattering rate, taking into account
uncertainties in the laser beam waist $60(10) \mum$, laser saturation uncertainty of 15\,\%,
and laser detuning uncertainty of $30 \MHz$. The overall uncertainty results in 15\,\% errors
for the excitation energy. The theoretical prediction of the simple model (dashed line)
for solely the error function transport does not agree with the data. However, we found
a motional excitation of $0.427(73)\meV$ due to the morphing steps of the potential
before and after the transport by replacing steps (iii) and (iv) in the measurement
cycle by a waiting time of equal length. The solid line shows this modified prediction
together with its standard deviation (dashed).} \label{fig:transportheating}
\end{minipage}
\end{figure}

After a transport, in step (vi) of the experimental sequence, the oscillating ion is
Doppler-cooled by laser light, and scatters an increasing amount of photons as it cools
down. We observe the scattered photons by a PMT as a function of  time. The scattering
rate in our model depends on the laser power quantified by a saturation parameter $s$,
the laser detuning $\Delta$, and the motional energy of the ion. In contrary to
\cite{WESENBERG2007} we do not average over a thermal state. For large excitations the
oscillation amplitude exceeds the waist of the cooling laser resulting in a low
scattering rate. A uniform laser intensity is therefore not appropriate for describing
our experiments. We take this effect into account by including a Gaussian beam waist of
$w_0 = 60\mum$ in our simulations. The efficiency of the laser cooling sets in with a
sharp rise of PMT counts shortly after $t_\text{recover}$. Only then, the scattering rate reaches its
steady state value at the Doppler-cooling limit, see
fig.~\ref{fig:transportheating}a. Adapting the theoretical treatment for a thermal
motional state in ref.~\cite{WESENBERG2007} to the case of a coherent motional
oscillation \cite{REICHLE200x} and including a spatial laser beam profile, the recovery
time of fluorescence is quantitatively identified with the energy after the transport.
The results are shown in fig.~\ref{fig:transportheating}b and compared to the
theoretical simulations of our simple classical model. As expected, the motional
excitation increases  for short transport times $\tau$. For slow transports, i.e.
$\tau\gg 1$, the excitation does not drop below a certain threshold. We found that this
excess excitation is due to the morphing steps (ii) and (v) and might be caused by a
non-ideal matching of the respective potential minima. Even a few $\mu$m difference are
leading to a large kick during the linear morphing ramp. We measured this effect
independently by replacing steps (iii) and (iv) by equal waiting times without voltage
ramps. Corrected for this excess heating, the model agrees well with the measured data.

The motional heating rate omitting any transport or morphing steps has been measured
independently for our trap by replacing steps (ii) to (iv) by waiting times of $500
\millis$ and $2000 \millis$. We deduce an energy gain of
$3(1)\,\text{meV/s}$. During our transport cycles this increase in energy only amounts to about
$1.5\,\mu\text{eV}$. This minor energy gain does not affect our measurement results
and conclusions on the transport-induced excitation.

\section{Conclusion and Outlook \label{outlook}}
Employing a novel segmented linear Paul trap we demonstrated stable trapping of single
ions and cold ion crystals. For the first time we have shown fast, non-adiabatic transports over $2\times 2\mm$ travels within a few microseconds by error function ramps. Main achievement is the
characterization of the ion's motional excitation. The method is based on the measured
modification of the ion's scattering rate during Doppler cooling. In the future,
however, sideband cooling and spectroscopic sideband analysis will be applied yielding
a much more sensitive tool to investigate motional energy transfers. This will lead to
a largely improved determination of ion excess motion. Then subtle changes of the
parameters and the application of optimal control theory for the voltage ramps may be
applied to yield lower motional excitation. Furthermore, we will establish a dedicated
loading zone and optimize the transport of single ions and linear crystals through the
tapered zone.

For micro-structured segmented ion traps with axial trap frequencies of several MHz,
the application of the demonstrated techniques will lead to transport times of a few
microseconds only. This way, future quantum algorithms may no longer be limited by the ion
transport but only by the time required for logic gate operations.

We acknowledge financial support by the Landesstiftung Baden-W\"urttemberg, the
Deutsche  Forschungsgemeinschaft within the SFB-TR21, the European Commission and
the European Union under the contract Contract No. MRTN-CT-2006-035369 (EMALI).

\vspace{1cm}
\bibliographystyle{unsrt}
%\bibliography{lit,techref}

\begin{thebibliography}{}

\end{thebibliography}


\begin{thebibliography}{10}

\bibitem{WINELAND2005}
D.~J. Wineland, et al., Proc. 17th Int. Conf. Laser Spec., World Scientific, 393, (2005).

\bibitem{BECHER2005}
C. Becher et al., Proc. 17th Int. Conf. Laser Spec., World Scientific, 381, (2005).

\bibitem{BARRETT2004}
M. D. Barrett, et al., Nature {\bf 429}, 737 (2004).

\bibitem{RIEBE2004}
M. Riebe, et al., Nature {\bf 429}, 734 (2004).

\bibitem{SCHMIDT2005}
P.O. Schmidt, et al., Science, {\bf 309}, 749, (2005).

\bibitem{ROSENBAND2007}
T. Rosenband, et al., Phys. Rev. Lett. {\bf 98}, 220801 (2007).

\bibitem{ROOS2006}
C. F. Roos, et al., Nature {\bf 443}, 316 (2006).

\bibitem{STICK2006}
D.~Stick, et al., Nature Physics {\bf 2}, 36, (2006).

\bibitem{SEIDELIN2006}
S.~Seidelin, et. al., Phys. Rev. Lett. {\bf 96} 253003, (2006).

%REV1
\bibitem{BROWN2007}
K. R. Brown, et al., Phys. Rev. A {\bf 75}, 015401,(2007).

\bibitem{GULDE01}
S.~Gulde, et al., Appl. Phys. B {\bf 73}, 861, (2001).

\bibitem{KIER00}
N. Kjaergaard, et al., Appl. Phys. B {\bf 71}, 207 (2000).

\bibitem{BERKELAND}
D.J. Berkeland,et al., J. Appl. Phys. {\bf 83}, 5025 (1998).

\bibitem{ROTTER2003}
D.~Rotter. {\em {Diploma thesis, Universit\"at Innsbruck}}, 2003.

\bibitem{SK2003}
F. Schmidt-Kaler, et al., Appl. Phys. B {\bf 77}, 789 (2003).

\bibitem{LEIBFRIED2007}
D.~Leibfried, E.~Knill, C.~Ospelkaus, and D.~J. Wineland, Phys. Rev. A {\bf 76}, 032324, (2007).

\bibitem{KIELPINSKI2002}
D.~Kielpinski, C.~Monroe, and D.J. Wineland, Nature, {\bf 417}, 709, (2002).

\bibitem{ChuangPC}
I.~Chuang, private communication.

\bibitem{HOME03}
J. P. Home at al., Quantum Inf. and Comput., {\bf 6}, 289, (2006).

\bibitem{HUCUL07}
D. Hucul et al., arXiv:quant-ph/0702175v2, (2007).

\bibitem{HENSINGER06}
W. K. Hensinger et al., Appl. Phys. Lett. {\bf 88}, 034101, (2006).

\bibitem{REICHLE2006}
R.~Reichle, et. al., Fortschr. Phys., {\bf 54}, 666, (2006).

\bibitem{SCHULZ2006}
S.~Schulz, U.~Poschinger, K.~Singer, and F.~Schmidt-Kaler, Fortschr. Phys., {\bf 54}, 648, (2006).

\bibitem{ROWE2002}
M.A. Rowe, et. al., Quantum Inf. and Comput., {\bf 2}, 257, (2002).

\bibitem{COUVERT07}
A. Couvert et al., arXiv:0708.4197v1, (2007)

\bibitem{WESENBERG2007}
J.H. Wesenberg, et. al., arXiv:quant-ph/0707.1314, (2007).

\bibitem{REICHLE200x}
R.~Reichle, et al., in preparation.

\end{thebibliography}

\end{document}